\documentstyle[twocolumn,prl,aps]{revtex}
\begin{document}                                                              
\draft
\title{Anomalous mesonic interactions near a chiral phase transition}
\author{Robert D. Pisarski}
\address{
Department of Physics, Brookhaven National Laboratory,
Upton, New York 11973-5000, USA}
\date{\today}
\maketitle
\begin{abstract}
Using constituent quarks coupled to a linear sigma model
at nonzero temperature, I show that many anomalous mesonic
amplitudes, such as $\pi^0 \rightarrow \gamma \gamma$,
vanish in a chirally
symmetric phase.  Processes which are 
allowed, such as $\pi^0 \sigma \rightarrow \gamma \gamma$,
are computed to leading order in a loop expansion.
\end{abstract}
\pacs{BNL Preprint BNL-RP-955, July, 1995. }
\begin{narrowtext}

The axial anomaly is the observation that for fermions
coupled to a gauge field,
the divergence of the current for axial fermion number 
is not just the standard contribution from the classical
equations of motion.  In addition, at one loop order the
divergence also contains a new term
from quantum effects \cite{anom1}.
Due to deep geometrical reasons, there are no further
corrections to this new term beyond one loop order \cite{geom}.  
For similar reasons, the axial anomaly is not altered by the 
presence of a medium, such
as a thermal bath, in either two \cite{anom3} or four
\cite{anom4} spacetime dimensions.

Besides the axial anomaly for fermions, there are also 
``anomalous'' mesonic interactions \cite{anom1,wzw}.  
These are like the fermion
anomaly, in that they arise from fermion loop graphs involving
an odd number of
the Dirac matrix $\gamma^5$, and so in the end are proportional
to the antisymmetric tensor $\epsilon^{\alpha \beta \delta \gamma}$.
Despite this superficial similarity, 
in this Letter I show that while the axial anomaly for fermions
is completely unaffected by the presence of a medium, 
near a chiral phase transition of second order,
anomalous mesonic interactions change dramatically.
As a medium I consider a thermal bath \cite{rdp3}, 
which may be produced in the central region of heavy ion
collisions at ultrarelativistic energies.

I begin by considering the prototypical 
anomalous mesonic interaction, the decay
of $\pi^0 \rightarrow \gamma \gamma$ \cite{anom1,wzw}.
Because the lifetime of the $\pi^0$ is 
electromagnetic, and so much longer than hadronic timescales, this
decay is not of experimental interest for heavy
ion collisions.  The essential physics, however, applies to the
(anomalous) decay of the $\omega$ meson, 
which does decay over hadronic timescales.

I work in a constituent quark model, with two
flavors and $N_c = 3$ colors of a quark field $\psi$.  
I couple quarks couple to photons, $A_\alpha$,
and to mesons, $\Phi$, ignoring their coupling to gluons.
I assume that the (quantum mechanical) breaking of the axial
$U(1)$ chiral symmetry is large at all temperatures \cite{pw}, so that 
the relevant chiral symmetry is
$SU(2)_\ell \times SU(2)_r$.
Then the only \cite{field} meson fields required are
$\Phi = \sigma t_0 + i \vec{\pi} \cdot \vec{t}$,
with $\sigma$ a $J^P=0^+$ meson, and $\vec{\pi}$ the $0^-$ pions;
the flavor matrices are $t_0 = {\bf 1}/2$ and
$tr(t^a t^b) = \delta^{a b}/2$.  

I take a positive definite euclidean metric, with $(\gamma^5)^2 = + 1$.
Left and right handed quark fields are constructed by
using the projectors $P_{\ell,r} = (1 \mp \gamma^5)/2$,
$\psi_{\ell,r} = P_{\ell,r} \psi$.  The lagrangian density 
for quarks is
\begin{equation}
{\cal L} =
\overline{\psi}_\ell  \not \!\! D  \psi_\ell
+ \overline{\psi}_r \not \!\! D \psi_r
+ 2 \widetilde{g} \left( 
\overline{\psi}_\ell \Phi \psi_r + 
\overline{\psi}_r \Phi^\dagger \psi_\ell \right) \; .
\label{eam}
\end{equation}
$\not \!\!\! D = (\not \! \partial - i q  \not \!\!\! A )$, 
where $q$ is a matrix
for the electric charge of the up and down quarks, 
$q = e (t_3 + t_0/3)$.  
Excluding the electromagnetic
coupling, this lagrangian is invariant under
global $SU(2)_\ell \times SU(2)_r$ chiral rotations $\Omega_\ell$
and $\Omega_r$, where
$\psi_{\ell,r} \rightarrow \Omega^\dagger_{\ell,r} \psi_{\ell,r}$
and
$\Phi \rightarrow \Omega_\ell^\dagger \Phi \Omega_r$.
With electromagnetism, $e \neq 0$,
the lagrangian is invariant under rotations
in the isospin-$3$ direction.  Explicitly,
\begin{equation}
{\cal L} =
\overline{\psi} \left( \not \!\! D + 2 \widetilde{g} 
\left( \sigma t_0 + i 
\vec{\pi} \cdot \vec{t} \gamma^5 \right)
\right) \psi \; .
\label{ea}
\end{equation}
If chiral symmetry breaking occurs, so
$\langle \sigma \rangle = \sigma_0$, I shift
$\sigma \rightarrow \sigma_0 + \sigma$, and the
consituent quark mass is $m = \widetilde{g} \sigma_0$.
At tree level, 
$\sigma_0 = f_\pi = 93 \, MeV$ is the pion decay constant.  

I neglect the dyanamics of the scalar and quark fields
to derive the effective lagrangian between the
scalar and photon fields which is induced by 
integrating out the quarks at one loop order.
Of course in an asymptotically free theory, at very high
temperatures mesons don't matter, 
only the quarks and gluons.  Mesons are important
at low and intermediate temperatures.  In particular,
for a chiral symmetry of 
$SU(2)_\ell \times SU(2)_r$, the chiral phase
transition can be of second order \cite{pw} at
$T= T_\chi$, which for the
sake of argument I assume is the case.  
I work in a strict chiral limit, so that
$\sigma_0(T)$ and $f_\pi(T)$ vanish as $T \rightarrow T_\chi$.
The pions are massless when $T \leq T_\chi$;
the $\sigma$ meson, which is heavy at zero temperature,
is massless at $T_\chi$, so that it and the pions form the
appropriate $SU(2)_\ell \times SU(2)_r = O(4)$ multiplet.
For three or more flavors the chiral phase transition is typically
of first order \cite{pw}; the present analysis is then of interest
if the transition is weakly first order.

Computing the amplitude for $\pi^0 \rightarrow \gamma \gamma$ 
is standard at zero temperature.
Let the two photons be $A_\alpha(P_1)$ and $A_\beta(P_2)$, where
$P_1$ and $P_2$ are the four momenta.  There are two triangle
diagrams which contribute; after doing the Dirac algebra, one
diagram contributes
\begin{equation}
- i \frac{4 \widetilde{g} e^2 N_c}{3} \; m  \; I(P_1,P_2,m) 
\epsilon^{\alpha \beta \delta \gamma} P_1^\delta P_2^\gamma \; , 
\label{eb}
\end{equation}
where $\epsilon^{\alpha \beta \delta \gamma}$ is
the antisymmetric tensor, and
$I(P_1,P_2,m)$ is the loop integral
\begin{equation}
tr_K \frac{1}{(K^2 + m^2) ( (K + P_1)^2 + m^2) ((K- P_2)^2 + m^2)}
\; .
\label{ec}
\end{equation}
$tr_K = \int d^4 K/(2 \pi)^4$ 
is the integral over the loop momentum $K$.
In the limit of small
momenta the dependence on $P_1$ and $P_2$ can be
neglected, with
%
%$I(0,0,m) = 1/(32 \pi^2 m^2)$.
\begin{equation}
tr_K \frac{1}{(K^2 + m^2)^3} = 
\frac{1}{32 \pi^2 m^2} \; .
\label{ed}
\end{equation}
The second diagram, which follows by interchanging
$P_1$ and $P_2$, and $\alpha$ and $\beta$, contributes equally.
Altogether, after using $m = \widetilde{g} f_\pi$,
the lagrangian density for
$\pi^0 \rightarrow \gamma \gamma$ is \cite{anom1,wzw}
\begin{equation}
i  \frac{e^2 N_c}{96 \pi^2 f_\pi}
\pi^0 
\epsilon^{\alpha \beta \delta \gamma}
F_{\alpha \beta} F_{\delta \gamma} \; .
\label{eda}
\end{equation}

To compute the corresponding amplitude at nonzero temperature I make
several assumptions.  First, I compute near the chiral
phase transition, $T\sim T_\chi$.
Since $\sigma_0(T) \rightarrow 0$, I can take
$m(T) = \widetilde{g} \sigma_0(T) \ll T$.  
Secondly, I work in the static limit, taking both
external energies to vanish,
$p_1^0 = p_2^0 = 0$.  Because of the antisymmetric tensor,
this means that implicitly I am assuming that one of the photons is a
plasmon, say $\alpha = 0$, while the other is spatial, $\beta = i$.
I also assume that the spatial momenta, $|p_1|$ and $|p_2|$,
are much smaller than the temperature.
These assumptions can be relaxed, but suffice to illustrate
how anomalous mesonic interactions change near $T_\chi$.

Under these assumptions, the computation of $\pi^0 \rightarrow \gamma\gamma$
at nonzero temperature
is utterly trivial.  To leading order in
$m$, all I have to do is to compute the integral
in (\ref{ec}) for $m=0$ and $T \neq 0$.
At $T \neq 0$, the fermionic loop momentum
$k^0 = (2n+1) \pi T$, summing over all integers $n$:
$$
tr_K \frac{1}{(K^2)^3} = 
T \sum_{n = - \infty}^{+ \infty}  \int
\frac{d^3 k}{(2 \pi)^3} 
\frac{1}{(k^2 + (k^0)^2)^3} $$
\begin{equation}
= \frac{1}{16 \pi ^4 T^2} 
\sum_{n = 1}^{\infty} \frac{1}{(2 n - 1)^3}
= \frac{7}{128} \; \frac{ \zeta(3)}{\pi^4 T^2} \; .
\label{ee}
\end{equation}
In doing the integral it is most convenient to first
integrate over the spatial momenta, and then do the sum over
the integers $n$.  This sum generates a zeta function,
$\zeta(r) = \sum^\infty_{n = 1} 1/n^r$;
in (\ref{ee}), $\zeta(3) = 1.20206...$ enters.
Consequently, in the static limit about $T_\chi$, 
after integration by parts 
the lagrangian density for $\pi^0 \rightarrow \gamma \gamma$ is
\begin{equation}
i \frac{7 \zeta(3) e^2 \widetilde{g}^2 N_c }{96 \pi^4 T^2}
\epsilon^{i j k}
\left(\sigma_0 \partial_i \pi^0 \right)
A_0 \partial_j A_k \; .
\label{ef}
\end{equation}

I emphasize that (\ref{ef}) holds only under
the given approximations \cite{pio,wzwt}, in particular, 
in the chiral limit {\it near} $T_\chi$.
At zero temperature (\ref{eda}) can be derived
using the partial conservation of the axial current ($PCAC$) and
the standard axial anomaly for fermions.  At low temperatures,
presumably something like (\ref{eda}) can by derived by using
$PCAC$ at $T\neq 0$;
at the very least, the zero temperature $f_\pi$ should be
replaced by $f_\pi(T)$ \cite{wzwt}.
In contrast, (\ref{ef}) is valid solely about
$T_\chi$, where $f_\pi(T)$ is small, and $PCAC$ breaks down.

It is natural for the constants in
(\ref{ef}) to differ from those in (\ref{eda}).
If they didn't, and the amplitude for 
$\pi^0 \rightarrow \gamma \gamma$ were proportional to $1/f_\pi(T)$
at all temperatures, 
then the amplitude would diverge as $T \rightarrow T_\chi$, 
when $f_\pi(T) \rightarrow 0$.  
Instead, the constants change from
$\sim 1/f_\pi$ at zero temperature
to $\sim \widetilde{g}^2 \sigma_0(T)/T^2$ near $T_\chi$,
so that instead of diverging, the amplitude 
vanishes as $T\rightarrow T_\chi^-$ \cite{second}.

I have written (\ref{ef}) in a suggestive manner.
The axial current for the scalar fields is
$\vec{J}_{A,i} = ( \sigma_0 + \sigma) 
\partial_i \vec{\pi} - \vec{\pi} \partial_i \sigma$.
%\begin{equation}
%\vec{J}_{A,\alpha} = ( \sigma_0 + \sigma) 
%\partial_\alpha \vec{\pi} - \vec{\pi} \partial_\alpha \sigma \; .
%\end{equation}
%
Thus (\ref{ef}) is part of the lagrangian density
\begin{equation}
i \frac{7 \zeta(3) e^2 \widetilde{g}^2 N_c}{96 \pi^4 T^2}
\epsilon^{i j k}
J^3_{A,i} A_0 \partial_j A_k \; .
\label{eg}
\end{equation}

To demonstrate that (\ref{eg}) is correct, observe that 
it predicts $\pi^0 \sigma \rightarrow \gamma \gamma$ even when
$\sigma_0 = 0$.  This process is given by six box diagrams, and
can be checked directly.  It is easiest to calculate the
box diagrams in two limits: first, when the $\sigma$ has zero momentum,
and then, when the $\pi$ has zero momentum.  The two coefficients
have the value given by (\ref{eg}), with the appropriate change
in sign.

The change in anomalous mesonic interactions near $T_\chi$ 
can be understood generally.  The restoration of chiral symmetry
at $T \geq T_\chi$ requires all couplings to be manifestly
chirally symmetric.  Hence electromagnetic amplitudes
must commute not just with the third component of the
vector charge (which is just isospin symmetry), but with
the third component of the axial vector charge as well.
This implies that 
$\pi^0 \rightarrow \gamma \gamma$ vanishes, and
constrains the amplitude for 
$\pi^0 \sigma \rightarrow \gamma \gamma$ to have the form
in (\ref{eg}), with a coupling directly to 
the third component of the axial current, $J^3_{A,i}$.
While the structure of the operator for
$\pi^0 \sigma \rightarrow \gamma \gamma$ 
is dictated by the chiral
symmetry, the coefficient in front is not; since it is
proportional to the coupling constant $\widetilde{g}^2$, 
whose value is arbitrary, it is not universal, and 
likely receives corrections in 
$\widetilde{g}^2$ from graphs to higher loop order.

Because I have computed in the static limit, the expression 
in (\ref{eg}) is gauge invariant: both
$\epsilon^{i j k} \partial_j A_k$ and $A_0$ 
are each unchanged under static gauge
transformations.  The amplitudes can also be computed
away from the static limit: then each external energy $p_0$
is analytically continued as $p_0 = - i \omega + 0^+$, where
$\omega$ is a continuous, minkowski energy.
For nonzero $\omega$ and $p$ the results
are more involved, and are nonlocal in coordinate space.  
In the limit of small momenta, where each $\omega$ and $p \ll T$,
and to leading order in $\widetilde{g}$ near $T_\chi$,
the resulting amplitudes are similar 
similar to those which arise for hard thermal loops \cite{htl}.
This is because the diagrams are at one loop,
with discontinuities due to
massless (fermion) fields at nonzero temperature.
In the end, the result for $\pi^0 \sigma \rightarrow \gamma \gamma$
must be gauge invariant, but in a more involved fashion.

With this example in hand, we can compute how many other
anomalous mesonic interactions change near $T_\chi$.
An example of importance for heavy ion collisions
is the $\omega$ meson.  The $\omega$ meson is special,
in that because it couples to the isosinglet
current for quark number, it only couples through
anomalous interactions \cite{ulf,rdp2}.
There are two anomalous interactions 
of importance at zero temperature, $\omega \rightarrow \rho \pi$ 
and $\omega \rightarrow \pi \pi \pi$.  
The corresponding lifetime of the $\omega$ meson is $\sim 20 fm/c$, which,
while long, is still of hadronic timescales.

To couple $\Phi$ to vector mesons, 
I assume strict vector meson dominance \cite{rdp2}.
Neglecting electromagnetism,
the covariant derivative in (\ref{ea}) becomes
\begin{equation}
\not \!\! D = \not \! \partial
-ig \left( \not \! \omega t_0 +
\not \! \vec{\rho} \cdot \vec{t} +
\gamma^5 \not \! \vec{a}_1 \cdot \vec{t} \; \right) \; ,
\end{equation}
where $g$ is the coupling constant to vector mesons.
Under chiral rotations the
isotriplets (the $J^P =1^-$ $\vec{\rho}$
and the $1^+$ $\vec{a}_1$) mix with each other, while the
isosinglets (the $1^-$ $\omega$ and the $1^+$ isosinglet $f_1$, which I
neglect) are invariant.

Up to trivial factors of isospin,
the diagrams which contribute to 
$\omega \rightarrow \rho \pi$ are the same as for
$\pi^0 \rightarrow \gamma \gamma$.  
To one loop order at zero temperature,
about zero momentum
the lagrangian density for $\omega \rightarrow \rho \pi$ is
\begin{equation}
- i \frac{g^2 N_c}{8 \pi^2 f_\pi}
\epsilon^{\alpha \beta \delta \gamma}
\omega_\alpha \partial_\beta \vec{\pi} 
\cdot \partial_\delta \vec{\rho}_\gamma \; .
\label{eh}
\end{equation}
Near the chiral phase transition, 
and taking a plasmonic $\omega_\alpha
= \omega_0$ to conform to 
the static limit, this becomes
\begin{equation}
- i \frac{ 7 \zeta(3) g^2 \widetilde{g}^2 N_c}{32 \pi^4 T^2}
\epsilon^{i j k}
\omega_0 \left(\vec{J}_{A,i} \cdot
\partial_j  \vec{\rho}_k
- \vec{J}_{V,i} \cdot \partial_j \vec{a}_{1,k} \right) 
\; ,
\label{ei}
\end{equation}
where the isospin vector current 
$\vec{J}_{V,i} = \vec{\pi} \times \partial_i \vec{\pi}$.
This expression is the product of a left handed current times a
left handed gauge field, minus the same for right handed quantities,
and so is invariant under 
the symmetries of chiral
$SU(2)_\ell \times SU(2)_r$ and parity.

For the second process, $\omega \rightarrow \pi \pi \pi$,
about zero momentum the lagrangian density is
\begin{equation}
+ i \frac{g N_c}{24 \pi^2 f_\pi^3} 
\epsilon^{\alpha \beta \delta \gamma} 
\omega_\alpha \partial_\beta \vec{\pi}
\cdot \partial_\delta \vec{\pi} \times
\partial_\gamma \vec{\pi} \; ,
\label{ej}
\end{equation}
to one loop order at zero temperature.
To compute the corresponding amplitude about the chiral phase
transition requires the integral
\begin{equation}
tr_K \frac{1}{(K^2)^4} = 
\frac{31}{1024} \; \frac{\zeta(5)}{\pi^6 T^4} \; ,
\label{eja}
\end{equation}
$\zeta(5) = 1.03693...$.  In the static limit near $T_\chi$,
the lagrangian density for a plasmonic $\omega_0$ is
\begin{equation}
i \frac{31 \zeta(5) g \widetilde{g}^4 N_c}{256 \pi^6 T^4}
\epsilon^{i j k}
\omega_0  (\sigma_0 + \sigma)  \partial_i \vec{\pi}
\cdot \partial_j \vec{\pi} \times
\partial_k \vec{\pi} \; .
\label{ek}
\end{equation}
Remembering that $\omega_0$ is a chiral singlet,
this can be written in a 
form which is manifestly chirally symmetric:
\begin{equation}
i \frac{31 \zeta(5) g \widetilde{g}^4 N_c}{64 \pi^6 T^4}
 \epsilon^{i j k}
\omega_0 \, tr \left( \Phi^\dagger \partial_i \Phi
\partial_j \Phi^\dagger  \partial_k \Phi
- ( \Phi \leftrightarrow \Phi^\dagger ) \right) \; .
\label{el}
\end{equation}

The anomalous interactions of the $\omega$ meson 
have dramatic consequences for 
a ``thermal'' $\omega$.
Without anomalous interactions, strict vector meson
dominance implies that the mass of the thermal
$\omega$ doesn't change with temperature \cite{rdp2}.
Including anomalous interactions, both the mass 
and the width of the thermal
$\omega$ {\bf must} change.  The present results demonstrate
that even the form of the anomalous interactions
change near $T_\chi$.
These changes may be observable by measuring a shift in
the $\omega$ peak, as seen in the 
dilepton spectrum in heavy ion collisions at ultrarelativistic
energies.  

I conclude by discussing one last anomalous process.
Extending the model to three flavors,
$\pi = \pi^a \lambda^a$ for the $SU(3)$ flavor matrices $\lambda^a$,
at zero temperature the lagrangian density for $KK \rightarrow \pi \pi \pi$
is \cite{wzw}
\begin{equation}
i \frac{2 N_c}{15 \pi^2 f_\pi^5} 
\epsilon^{\alpha \beta \delta \gamma} 
tr\left( \pi \partial_\alpha \pi \partial_\beta \pi
\partial_\delta \pi \partial_\gamma \pi \right) \; .
\label{em}
\end{equation}
Using the integral
\begin{equation}
tr_K \frac{1}{(K^2)^5} =
\frac{635}{32768} \; \frac{\zeta(7)}{\pi^8 T^6} \; ,
\label{ema}
\end{equation}
where $\zeta(7) = 1.00835...$,
then about the chiral phase transition, to one loop order the
lagrangian density between one $\sigma$ meson with zero momentum and
five $\pi$'s is
\begin{equation}
i  \frac{127 \sqrt{2} \zeta(7) \widetilde{g}^6 N_c}{256 \sqrt{3}\pi^8 T^6}
\epsilon^{i j k} (\sigma_0 + \sigma)
tr\left( \pi \partial_0 \pi \partial_i \pi
\partial_j \pi \partial_k \pi \right) \; .
\label{en}
\end{equation}
As before, (\ref{en}) implies that while the amplitude for
$KK \rightarrow \pi \pi \pi$ vanishes in a chirally symmetric
phase, that for $KK \rightarrow \pi \pi \pi \sigma$ does not.
While I have written (\ref{en}) in a form which appears to be
Lorentz invariant, it is valid {\it only} in the limit where 
$\sim \partial_0 \pi$ is vanishingly small.  Terms
of higher order in the frequency involves expressions which
are nonlocal in coordinate space.

At zero temperature (\ref{em}) is the first term in the expansion
of a chirally symmetric lagrangian in five dimensions.
Witten showed that
the form of this lagrangian is dictated by topology in 
five dimensions \cite{wzw},
which fixes the constant in front, up to an overall integer equal
to the number of colors.
Likewise, it must be possible to write (\ref{en}) in chirally
invariant form.  Note, however, that the constant
in front of (\ref{en}) does not appear to be constrained
by topology, since it involves a coupling constant, $\widetilde{g}$,
whose value is arbitrary \cite{topology}.

The following picture emerges.  At low temperatures,
only pions are massless, with their anomalous
interactions governed by the generalization
of the Wess-Zumino-Witten lagrangian \cite{wzw}
to nonzero temperature \cite{wzwt}.  
If the chiral transition is of second (or weakly first) order,
though, then near $T_\chi$ anomalous mesonic interactions
are governed by a new lagrangian, which includes the
terms in (\ref{eg}) and (\ref{en}) as two examples.
A new lagrangian emerges because
near the critical point new modes ---
the $\sigma$ mesons \cite{wzws} --- become light,
so the anomalous mesonic lagrangian adjusts to include them.

I thank S. Aronson for emphasizing the experimental 
importance of the $\omega$ peak, which led me to consider
this problem.  I also thank A. Bochkarev, M. Creutz, S. Dawson,
S. Gavin, A. Kovner, M. Mattis, M. Tytgat, M. Shifman, L. Trueman, 
M. Voloshin, and A. Weldon for discussions and comments.
After this work was completed, I received \cite{baier};
based upon \cite{rdp3}, \cite{baier} obtains results similar
to those in this Letter.
This work is supported by a DOE grant at 
Brookhaven National Laboratory, DE-AC02-76CH00016.

\end{narrowtext}

\begin{references}
\bibitem{field}
If the axial $U(1)$ symmetry is approximately restored
before the chiral phase transition \cite{pw}, then it is
also necessary to introduce the $J^P=0^-$ isosinglet $\eta$ meson and
an isotriplet of $0^+$ $\vec{\sigma}$ mesons.  For three or more 
flavors, $n_f \geq 3$, regardless of what happens to the axial
$U(1)$ symmetry, $\Phi$ must include
not just $\sigma$ and the $n_f^2 -1 $ $\pi^a$'s, 
but the $\eta'$ and the $n^2_f - 1$ $\sigma^a$'s as well.
This is because under
$SU(n_f)_\ell \times SU(n_f)_r$ rotations,
not only does the $\sigma$ mix with the $\pi^a$'s,
and the $\eta'$ with the $\sigma^a$'s, but, 
due to the symmetric flavor tensor $d^{a b c}$,
the $\sigma^a$'s mix with the $\pi^a$'s.
\bibitem{second}
Near $T_\chi$, the
amplitude for $\pi^0 \rightarrow \gamma \gamma$, 
like the analogous pieces in (\ref{ei}), (\ref{ek}), and (\ref{en}), 
vanishes $\sim \sigma_0^\xi$ as $\sigma_0 \rightarrow 0$,
with $\xi = 1$.  Perhaps, because of the presence of the
antisymmetric tensor, this mean field value of $\xi$ is exact.
\bibitem{topology}
The constraints of topology can be subtle at nonzero temperature.
For example, in topologically massive gauge theories
in three spacetime dimensions, classically topology restricts
the ratio of the topological mass to the
coupling constant to be an integer
\cite{djt}.  To one loop order, perturbative calculation of the
Green's functions at small momentum show that 
in $SU(N_c)$ gauge theories, this integer is shifted by
$+N_c$ \cite{pr}.  At nonzero temperature,
the topological quantization condition still applies,
but topology doesn't restrict the form of the Green's function at small
momentum \cite{rdp5}, as it does at zero temperature.
\bibitem{wzws}
For three or more flavors, $n_f \geq 3$,
$\Phi$ must include all $2 n_f^2$ fields \cite{field}.
Besides interactions 
$\sim \sigma (\pi^a)^5$ as in (\ref{en}), 
there are many others, including
$\sim \sigma (\sigma^a)^2 (\pi^a)^3$,
$\sim \sigma (\sigma^a)^4 \pi^a$,
$\sim \eta' (\sigma^a)^5$, 
$\sim \eta' (\sigma^a)^3 (\pi^a)^2$,
and $\sim \eta' \sigma^a (\pi^a)^4$.
Like other anomalous mesonic interactions,
these terms are all odd under Witten's \cite{wzw} transformation
of $(-1)^{N_\pi}$, where
$\eta' \rightarrow - \eta'$, $\pi^a \rightarrow - \pi^a$.
\end{references}
\end{document}